\documentclass[twocolumn]{aastex701}
\usepackage{amsmath}

\begin{document}

\title{Photometry is all you need: supernova classification as a mixing problem}

\author[0000-0001-9308-0449]{Ana Sof{\'i}a M. Uzsoy}
\affiliation{Center for Astrophysics $|$ Harvard \& Smithsonian, 60 Garden St., Cambridge, MA 02138, USA}
\email[show]{ana\_sofia.uzsoy@cfa.harvard.edu}

\author[orcid=0000-0002-5814-4061]{V. Ashley Villar} 
\affiliation{Center for Astrophysics $|$ Harvard \& Smithsonian, 60 Garden St., Cambridge, MA 02138, USA}
\affiliation{The NSF AI Institute for Artificial Intelligence and Fundamental Interactions}
\email{ashleyvillar@cfa.harvard.edu}

\begin{abstract}

In the era of large-scale photometric surveys, scalable and robust methods for classifying supernova (SN) populations are increasingly necessary. Often, spectroscopy is essential in addition to photometry to reliably classify SNe; however, complete spectroscopic follow-up is infeasible for all of the millions of transient light curves being collected by facilities such as the Vera C. Rubin Observatory. Using light curves of SNe Ia and Ibc observed with the Zwicky Transient Facility, we frame the classification of large SN populations as a mixing problem. We fit all objects using a semi-analytical SN model powered by radioactive decay, and we model the resulting distributions of fit parameters with a Gaussian Mixture model to optimize the shared population mixing fraction. This approach allows us to reliably constrain the ratio of the populations and classify SNe Ia and Ibc with $\geq$ 90\% accuracy without any need for labeled training data, i.e., a spectroscopic dataset. We validate this method for varying population mixing fractions and explore the impact of including spectroscopic, photometric, or no redshift information, and a small amount of known labels. Overall, this method allows for fast and accurate SN classification and population characterization using only photometry.
\end{abstract}



\section{Introduction} 
Ground-based, survey telescopes such as the  Zwicky Transient Facility \citep[ZTF;][]{2019PASP..131a8002B} and the Vera C. Rubin Observatory \citep[LSST;][]{lsstsciencecollaboration2009lsstsciencebookversion} are revolutionizing the field of transient astronomy by aiming to observe millions of candidate supernovae (SNe). As a result, the need for scalable computational methods to perform SN classification and characterization is greater than ever. While spectroscopy provides critical information about an object's intrinsic physics \citep{filippenko_optical_1997,branch2017supernova}, the vast majority of transients ($>$90\%) will only ever be observed with photometry \citep{kulkarni2020integratedopticaltransientutility}.

Classification of observed SNe into subtypes is critical for downstream science tasks, including understanding the progenitors of core-collapse SNe (e.g., \citealt{das2025low}), evaluating the contributions of various engines to unique SN subtypes (e.g., \citealt{afsariardchi2021nickel}), and cosmological parameter estimation (e.g., \citealt{Brout_2022}). Significant work has been done to leverage machine learning techniques for SN classification using both real and simulated photometric light curves \citep{gomez_fleet_2020,boone2021parsnip, villar19classification, hosseinzadeh2020photometric, villar20superraenn, sanchessaez21alerce, de_soto_superphot_2024, shah2025oracle, boesky26}. Additionally, there has been an emphasis on scalable methods to characterize SNe that can be widely applied to large datasets \citep{villar2022amortized, uzsoy+24, grayling+24,vidal2025hierarchical}.


Creating a scalable, flexible classification model that be widely applied to transient datasets is nontrivial, namely because of domain shift and class imbalances. Domain shift denotes the problem of applying a model which has been created for use with one dataset to another dataset. Many SN classification models are trained only on data from one survey taken under a specific set of observational parameters and cannot be applied to data from a different survey \citep{2024sharing}. SN datasets are also often highly imbalanced, with SNe Ia typically comprising $\sim$70\% of all observed light curves \citep{perley2020zwicky}, which makes it difficult to create models that perform well across different SN types.  While these can be addressed with stratified training and testing sets, modified loss functions, fine-tuning to new data, and other approaches, they all require a ground-truth labeled training set tuned to the specifics of the dataset they are to be applied to, which often involves laborious visual inspection.

We present a new method that classifies transients as SN Ia or Ibc on a population level without the need to spectroscopically classify any individual object. We focus on these two classes as they are, photometrically, the most likely to be misclassified as each another due to their similar underlying heating source (radioactive decay). We follow the Bayesian Estimation Applied to Multiple Species \citep[BEAMS;][]{beams} framework, where we establish our goal as a population mixing problem, and we seek to constrain the ``contamination fraction" of SNe Ibc in a population of SNe Ia. By fitting each light curve with a semi-analytic model, we fit a Gaussian Mixture model (GMM) to the resulting distribution of fit parameters and constrain the population mixing fraction to characterize the two populations in the dataset. We examine the accuracy of our method under simulated spectroscopic, photometric, or no redshift constraints. We find that fits using simulated photometric redshifts do almost equally as well as those using spectroscopic redshifts, but the no-redshift case does much worse. Finally, we re-fit under the assumption that 10\% of the population has known (i.e., spectroscopic) classifications and find that it does not significantly improve accuracy for the photometric or spectroscopic redshift cases.

This paper is organized as follows: in Section~\ref{sec:methods} we describe the ZTF data used as well as the light curve model and our GMM approach to classification. In Section~\ref{sec:results} we present the results of our method under different redshift conditions for populations of varying SN Ibc/Ia ratios, and discuss these results in Section~\ref{sec:discussion} before concluding in Section~\ref{sec:conclusions}.

\section{Methods} \label{sec:methods}
\subsection{Data}

ZTF observes transients in optical wavelengths using the 48-inch Schmidt telescope at the Palomar Observatory \citep{bellm19ztf} and so far has observed well over 30,000 extragalactic transients with a typical cadence of three days \citep{rehemtulla_2024, wise_2024}. The Bright Transient Survey (BTS) selects transients in the northern sky with a peak magnitude less than 18.5 mag for spectroscopic classification \citep{fremling_zwicky_2020, perley2020zwicky} and has thus far classified $\sim7000$ SNe. 

We use 3703 SN Ia and 255 SN Ibc photometric light curves in the $g$ and $r$ bands from ZTF BTS that also have redshift and dust extinction information \citep{fremling_zwicky_2020, perley2020zwicky}. For the former, we use the sample of light curves presented by \cite{maven}. For the SNe Ibc, we query the BTS Sample Explorer through December 2024 for Ib/Ic/Ibc types, retrieve 261 events, and exclude 6 that are ``peculiar" types or have no spectroscopic redshift values.

We use data in the $g$ and $r$ bands (as $i$ band tends to be much more sparsely sampled), and we exclude any objects with less than 5 points in each band, leaving 197 SN Ibc and 2913 SN Ia light curves, and we retain the first 2500 SN Ia light curves. Some light curves that do not converge are removed from the sample (see Section~\ref{sec:lc}). All SNe in our sample have spectroscopic redshifts between $z = 0$ and $z = 0.15$. We minimally process the light curves, except that we align the light curves such that $t = 0$ occurs at the time of first detection (3$\sigma$).

SNe Ia are much more common observationally than SNe Ibc \citep{kessler_2019}, and thus most transient datasets contain disproportionately more SNe Ia than SNe Ibc. To simulate different Ibc/Ia ratios, we create 10 subsets of our total dataset with ratios ranging from 0.1 to 0.5. To create these subsets, we always use all converged SN Ibc light curves, but we include varying amount of SN Ia light curves. For the spectroscopic dataset, the 0.5 ratio dataset contains 193 SNe Ibc and 193 SNe Ia while the 0.1 ratio dataset contains 193 SNe Ibc and 1,737 SNe Ia.

\subsection{Light Curve Models}
To fit all light curves, we use the semi-analytic ``Arnett" Model \citep{arnett1982}, which models a SN as being powered by radioactive decay of $^{56}$Ni and $^{56}$Co. The luminosity is written as a function of time as follows (see Equation 9 in \citealt{chatzopoulos_generalized_2012}):
\begin{equation}
\begin{split}
    L(t) = \frac{2M_{\rm Ni}}{t_d}e^{-\frac{t^2}{t_d^2}}\biggl[ & (\epsilon_{\rm Ni} - \epsilon_{\rm Co}) \int_0^t \frac{t'}{t_d} e^{\left(\frac{t'^2}{t_d^2} - \frac{t'}{t_{\rm Ni}}\right)}dt' \\
    &+ \epsilon_{\rm Co} \int_0^t \frac{t'}{t_d} e^{\left(\frac{t'^2}{t_d^2}- \frac{t'}{t_{\rm Co}}\right)}dt'\biggr]
\end{split}
\end{equation}
where $M_{\rm Ni} = f_{\rm Ni} M_{\rm ej}$, or the product of the ejecta mass ($M_{\rm ej}$) and the fraction of ejecta that is initially $^{56}$Ni ($f_{\rm Ni}$); $t_{\rm Ni}$ is the characteristic decay time of $^{56}$Ni and set to 8.8 days; $t_{\rm Co}$ is the characteristic decay time of $^{56}$Co and set to 111.3 days; $\epsilon_{\rm Ni} = 3.9 \times 10^{10}$ erg g$^{-1}$ s$^{-1}$ and $\epsilon_{\rm Co} = 6.8 \times 10^{9}$ erg g$^{-1}$ s$^{-1}$, denoting the energy generation per unit mass and time from Ni and Co, respectively; and $t_d$ is the effective diffusion timescale:
\begin{equation}
    t_d = \sqrt{\frac{2 \kappa M_{\rm ej}}{13.7 c v_{\rm ej}}}
\end{equation}
where $\kappa = 0.1$cm$^2$ g$^{-1}$ is the opacity, $c$ is the speed of light and $v_{\rm ej}$ is the characteristic ejecta velocity.

We also explored fitting all SNe using the SALT2 SN Ia light curve model \citep{guy_salt2_2007} in addition to the Arnett fit. In practice, we find that it was challenging to fit all parameters simultaneously to a GMM, especially with a full-rank covariance matrix. In fact, we ultimately find that using just a subset of two parameters from the Arnett model ($\log_{\rm 10} v_{\rm ej}$ and $\log_{\rm 10} f_{\rm Ni}$), yield the best classification results. A full discussion of fitting all parameters and its challenges can be found in Appendix~\ref{appendix:fitting_all}.

\begin{figure*}
    \epsscale{1.2}
    \centering
    \plotone{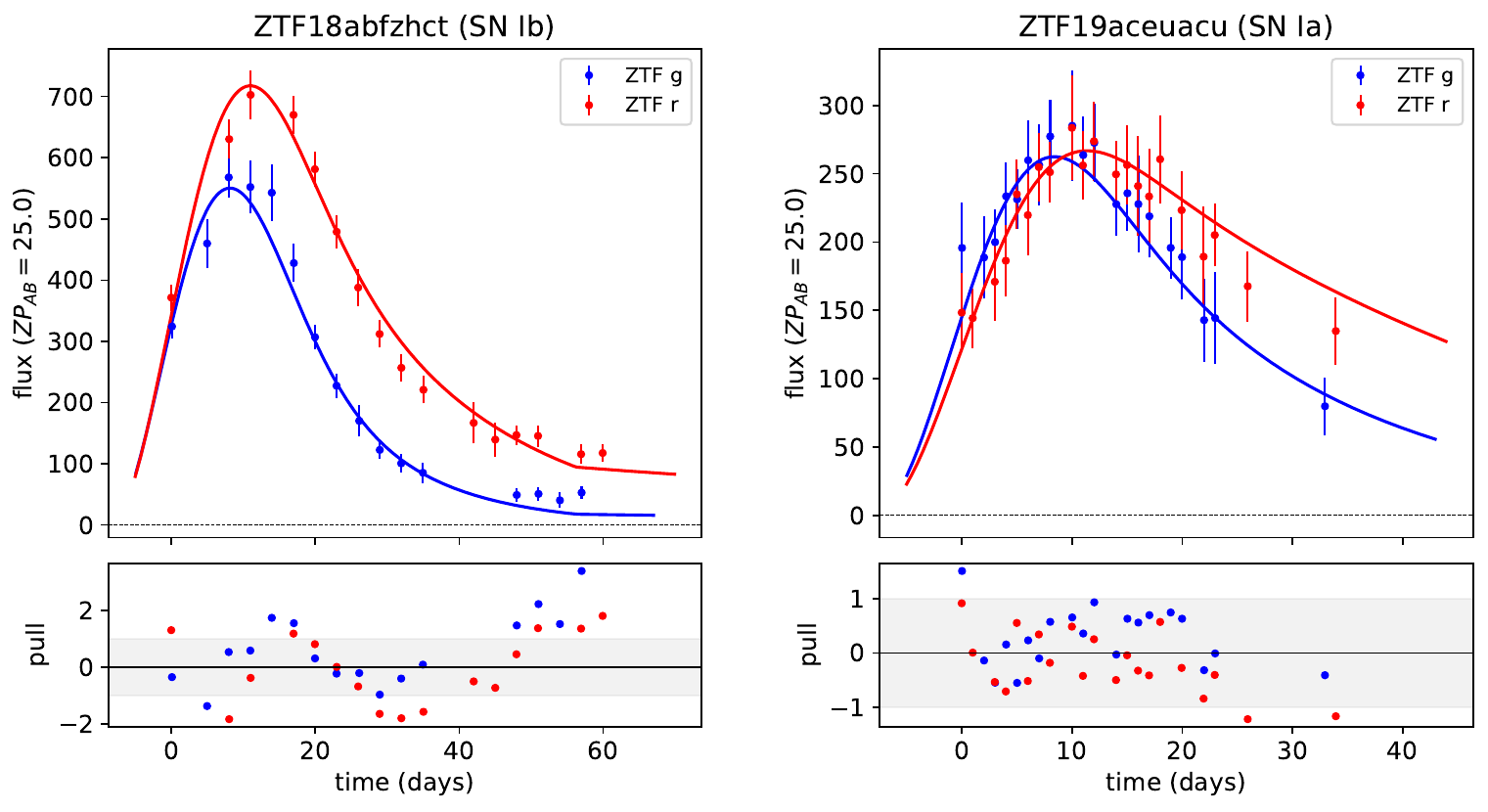}
    \caption{Example fits of SN Ia and Ibc light curves with the Arnett light curve model using {\tt sncosmo} for the ZTF $g$ band (blue) and $r$ band (red). Points indicate observed data points while solid lines indicate model fits. Bottom panels indicate the ``pull", (data - model)/error. Shaded regions indicate a pull between -1 and 1.}
    \label{fig:example_ibc_model_fit}
\end{figure*}

\subsection{Light curve fitting procedure}
\label{sec:lc}
We use the {\tt sncosmo} Python package \citep{barbary_2025_15019859} to fit the light curves from ZTF. We create a custom Arnett model by extending the built-in {\tt sncosmo.Source} class and fit all light curves with the {\tt sncosmo.fit\_lc} function, which optimizes parameters based on a $\chi^2$ minimization using the {\tt minuit} algorithm. We also account for dust extinction using the \citet{ccm89} dust map. Figure~\ref{fig:example_ibc_model_fit} shows sample light curve fits for a representative SN Ibc and SN Ia using the Arnett model, each with known (spectroscopic) redshifts. The Arnett model is initialized with [$t_{\rm exp}, m_{\rm ej}, f_{\rm Ni}, v_{\rm ej}$] = [0, 3, 0.05, 15000], and if it does not converge, the fit is re-run with an initialization of [10, 0.5, 0.1, 20000] to probe a different area of parameter space. Four, four, and five SN Ibc and 70, 84, and 77 SN Ia light curve fits did not converge for the spectroscopic, photometric, and no redshift cases, respectively, and are not included in the final sample.

\begin{figure*}
    \epsscale{1.15}
    \centering
    \plotone{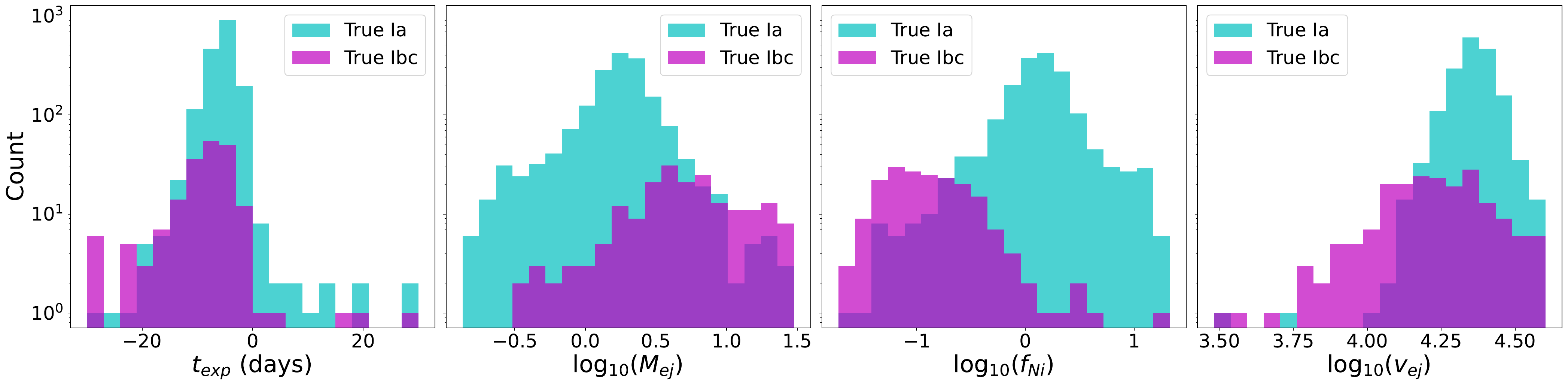}
    \caption{The distributions of fit parameters for the entire dataset, with 1737 SNe Ia in cyan and 193 SNe Ibc in magenta (with a Ibc fraction of 0.1), as SNe Ibc with the Arnett model. Y-axis is $\log_{10}$ scaled and shared for all parameters.}
    \label{fig:fitted_params}
\end{figure*}

Table~\ref{tab:arnett_params} shows the parameters we fit for each SN and their fitting bounds. The bounds are chosen to be broad to impose relatively uninformed and uniform priors on all parameters.  
Note that $f_{\rm Ni}$ should physically only take a value between 0 and 1; however, for the purposes of constraining the mixing fraction to separate populations, we allow it to reach nonphysical values greater than 1. Such a value may be due to, e.g., some additional heating source or engine not accounted for in our simple model. Figure~\ref{fig:fitted_params} shows the distribution of fit parameters for the Arnett model for all SNe in our sample.

\begin{table}[t]
\centering
\caption{Arnett light curve model parameters and fitting bounds.}
\label{tab:arnett_params}
\begin{tabular}{lcc}
\hline
Parameter & Lower Bound & Upper Bound \\
\hline
$v_{\rm ej}$ (km/s)    & 2,000 & 40,000 \\
$M_{\rm ej}$ ($M_\odot)$ & 0.01 & 30 \\
$f_{\rm Ni}$ & 0.01 & 50 \\
$t_{\rm exp}$ (days) & -30 & 30 \\
\hline
\end{tabular}
\end{table}

Additionally, we seek to investigate the effect of redshift uncertainties in our method. We fit each light curve three times: once under the assumption of a spectroscopic redshift estimate, once assuming a photometric redshift estimate, and once assuming no redshift information is available. For the ``spectroscopic redshift" case, we assume that the redshift is well-constrained and set it in the model as ground truth without fitting it. For the ``photometric redshift", the redshift is fit as a parameter in the model with broad bounds $(0.5z_{\rm spec}, 1.5z_{\rm spec})$, where $z_{\rm spec}$ is the spectroscopic redshift. In the ``no redshift" case, the redshift is an additional parameter that is fit with the loose bounds of (0, 0.2).

\subsection{Gaussian Mixture Models (GMMs)}
A GMM characterizes a probability distribution as a weighted sum of Gaussian distributions. GMMs are parameterized by the mean vectors and covariance matrices of their constituent Gaussian distributions and the fractional weight of the contribution of each component to the total. The two-Gaussian case is parameterized by the mixing fraction $\alpha$, with one Gaussian weighted by $\alpha$ and the other by (1 - $\alpha$). In the case of fitting a GMM to a distribution of SNe belonging to two classes, the mixing fraction denotes the fraction of the population that belongs to one class. Our objective is to leverage the distributions of light curve fit parameters to constrain the population-level mixing fraction $\alpha$, derive probabilities of each object belonging to each class, and ultimately disentangle the two populations.

We express the distribution of the vector of our fitted parameters (or, ultimately, a subset of them) $\mathbf{x}$ as:
\begin{equation}
    \mathbf{x} \sim \alpha \mathcal{N}(\boldsymbol{\mu_{Ibc}}, \mathbf{\Sigma_{Ibc}}) + (1 - \alpha) \mathcal{N}(\boldsymbol{\mu_{Ia}}, \mathbf{\Sigma_{Ia}})
\end{equation}
where $\alpha$ is the mixing fraction of SNe Ibc and $\boldsymbol{\mu_{Ibc}}, \boldsymbol{\mu_{Ia}}, \mathbf{\Sigma_{Ibc}}$, and $ \mathbf{\Sigma_{Ia}}$ are the mean vectors and covariance matrices of the two multivariate Gaussian distributions optimized to fit the two components of parameter vector $\mathbf{x}$. In taking this approach, we are assuming that there are only two populations in our dataset and that the distributions of fit parameters for each population are approximately Gaussian.

We use the Expectation-Maximization (EM) algorithm \citep{em_algorithm} to optimize the parameters of the GMM. The EM algorithm alternately iterates to optimize the means and covariances of the constituent Gaussian distributions and the value of the mixing fraction to best fit the data. The expectation step (E-step) calculates weights $\gamma_{Ia}$ and $\gamma_{Ibc}$ for each object, which denote the likelihood of it belonging to each class (and by definition sum to 1) under the current proposed Gaussian distributions. The maximization step (M-step) then re-estimates the means and covariances of the Gaussian distributions based on the class weights of each object to optimize the total log-likelihood of the full population. Once the fit is complete, the SN $s$ can be classified as SN Ia or Ibc by comparing the weights $\gamma_{Ia,s}$ and $\gamma_{Ibc,s}$. We optimize the parameters over 200 epochs to ensure that the log-likelihood is converged.

GMMs can be quite sensitive to their initialization values and are prone to converge to local minima. To ensure the best possible fit, we use 2D KMeans clustering to separate the data into potential classes, then initialize the  means and covariances from these classes before running the GMM fit. We repeat the fit 30 times, each with a different initialization, and we finally use the fit with the highest log-likelihood value. We initialize the mixing fraction $\alpha$ at 0.5 for all iterations.

The mixing fraction $\alpha$ is not necessarily constrained to be the fraction of SNe Ibc (i.e. either of the two components could correspond to SNe Ibc); however, we assume that the majority class corresponds to SNe Ia and use the component assignments that maximize accuracy. Note that we do not use the known labels to classify objects or constrain the mixing fraction; they are only used to evaluate accuracy or to anchor the GMM fitting in the ``10\% known" analysis (see Section~\ref{sec:results}).

Once the GMM fitting is completed, an individual SN $s$ is classified as SN Ia or Ibc using the value of $\gamma_{Ia,s}$, which denotes the probability of belonging to the Gaussian component corresponding to SNe Ia. We classify any object $s$ with $\gamma_{Ia,s} \geq 0.5$ as SN Ia, with all others being classified as SN Ibc. We then calculate the accuracy compared to the known labels from the dataset. In addition, we can compare the final mixing fraction value with the known population split in the dataset.

\subsection{``10\% known" analysis}

In many cases, photometry is obtained for a large number of transients, but a smaller fraction are selected for spectroscopic follow-up. Spectroscopy is considered the gold standard for characterizing SNe, and it is likely that obtaining a spectrum would lead to a definitive classification of an object as SN Ia or Ibc. Following the estimate of \cite{kulkarni2020integratedopticaltransientutility} that only 10\% of transients will have spectroscopy, we modify our analysis to simulate this case (hereafter referred to as the ``10\% known" analysis) and explore how much the classification accuracy on the entire population can be improved by incorporating a few known classifications. We note that we choose a $10\%$ subset at random, vs selecting e.g., the brightest 10\%, which may be more realistic. Given our training set is entirely from the spectroscopic BTS sample, this likely has a minimal effect in our results. Note that this sampled ``known" data is not stratified by class, but we would expect the ratio of known SNe Ia to Ibc to be roughly the same ratio as that of the entire dataset. 

In our GMM framework, each object $s$ has two weights $\gamma_{Ia,s}$ and $\gamma_{Ibc,s}$, which denote the probability of that object belonging to the component for SNe Ia or Ibc.
We first calculate, in the expectation step, the maximum value of $\gamma$ ($\gamma_{\rm max}$) across all objects and classes. For each ``known" object, we rescale the weight for the correct class to 10$\gamma_{\rm max}$ and set the weight for the incorrect class to zero. For example, if SN $s$ is classified as type Ia, then we set $\gamma_{Ia,s} = 10 \gamma_{\rm max}$ and $\gamma_{Ibc,s} = 0$. The weights for all objects are then renormalized to sum to one before proceeding with the maximization step.

In doing this reweighting, we are essentially forcing the SN Ia and Ibc distributions to incorporate their corresponding ``known" objects, which guides the fitting of their parameters. We initialize our distributions in the same way as the naive case, using 30 repetitions of KMeans, but each repetition uses a different randomly selected 10\% of the data as the ``known" labels. We ensure that the clusters identified by KMeans are aligned with the distribution of the known objects, swapping the components if necessary. After the GMM is fit, SNe are classified using the $\gamma_{Ia}$ weights as before.

\section{Results} \label{sec:results}
Here we explore the performance of this method on SN populations with different contamination fractions, redshift information, and known label fractions. 

\begin{figure}
    \epsscale{1.15}
    \centering
    \plotone{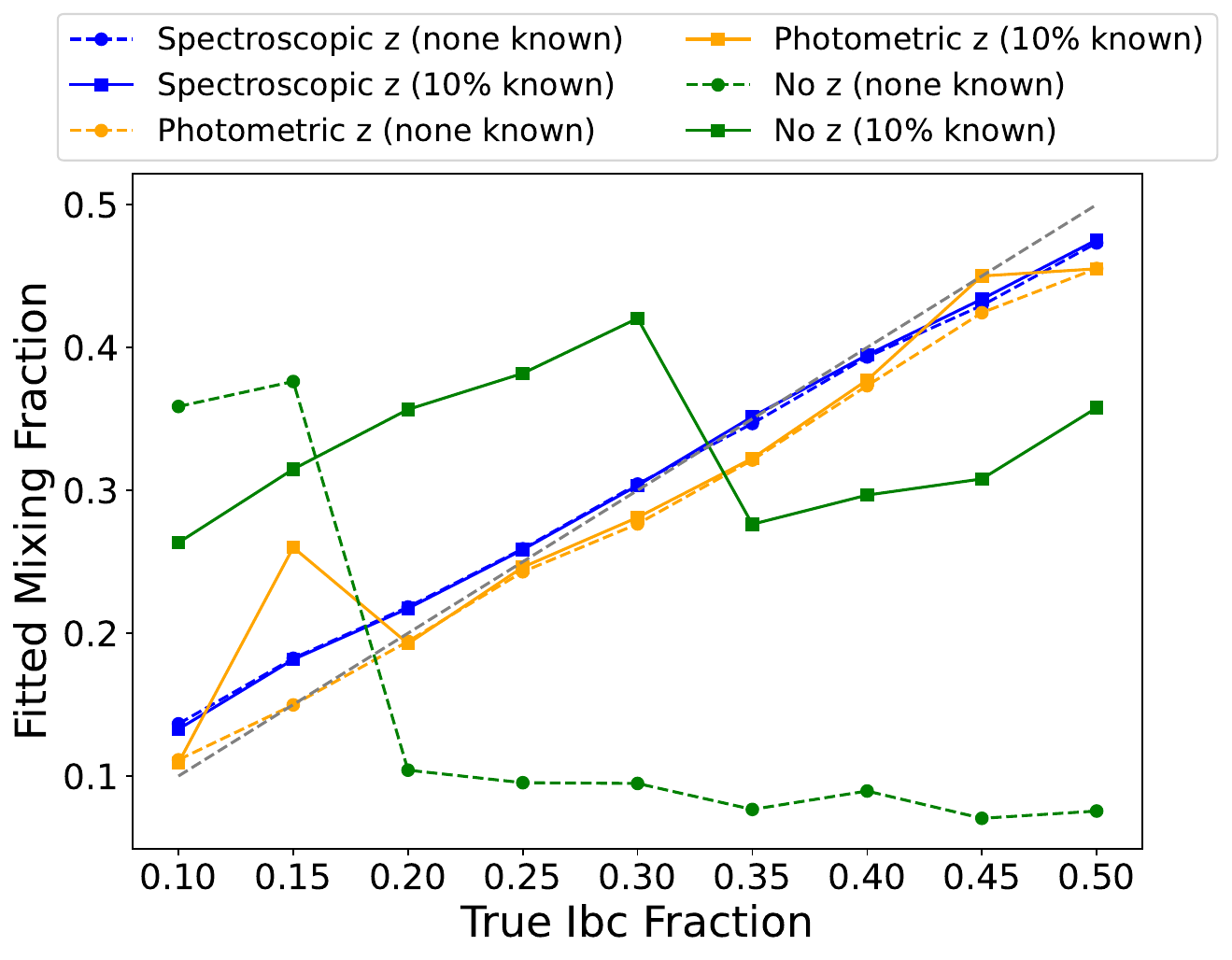}
    \caption{Fitted mixing fraction ($\alpha$) from our GMM fit based on the $\log_{10} f_{\rm Ni}$ and $\log_{10} v_{\rm ej}$ features as a function of the true SNe Ibc fraction in the population. Light curve fits were done using spectroscopic redshifts (blue), photometric redshift estimates (yellow), and no redshift information (green). Dashed lines denote accuracy assuming no labels are known, and solid lines denote accuracy with labels known for a randomly sampled 10\% of the dataset. Gray dashed line denotes 1:1 for comparison.}
    \label{fig:mixing_fraction_redshift}
\end{figure}

Figure~\ref{fig:mixing_fraction_redshift} shows the fitted vs. true Ibc/Ia mixing fraction for all redshift cases and with and without 10\% known. Both the spectroscopic and photometric redshift models fit mixing fractions close to the true value for all populations. At a true mixing fraction of 0.15, the known classifications include a small number of SNe Ibc that have high $f_{\rm Ni}$ and $v_{\rm ej}$ values, similar to those expected from true SNe Ia, which leads the model to overestimate the mixing fraction. However, this is a result of small-number statistics, and overall, inclusion of the 10\% known classifications makes little difference to the spectroscopic and photometric redshift models. In contrast, the no-redshift model with no known classifications performs poorly, more often than not heavily underestimating the mixing fraction. This is likely because the redshift is degenerate with other model parameters and our likelihood-optimization inference often leads to reasonable but incorrect fits (see Appendix~\ref{appendix:example_fits} for example fits using all three models). 
Unlike the spectroscopic and photometric redshift cases, the performance of the no-redshift case does improve with the inclusion of known classifications, although the overall performance is still significantly worse than than the other two cases.

\begin{figure}
    \epsscale{1.15}
    \centering
    \plotone{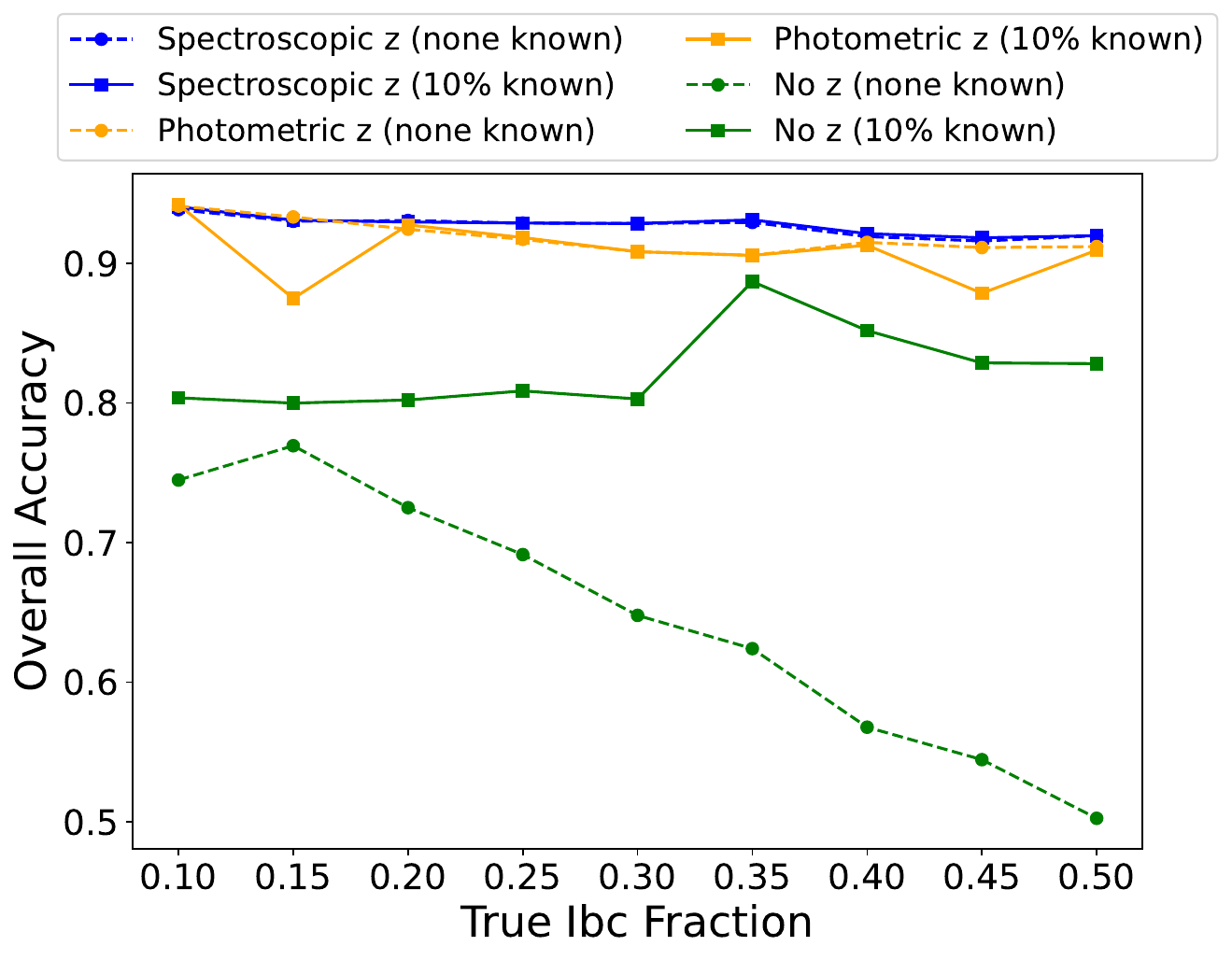}
    \caption{Overall classification accuracy our GMM fit based on the $\log_{10} f_{\rm Ni}$ and $\log_{10} v_{\rm ej}$ features as a function of the true SNe Ibc fraction in the population. Light curve fits were done using spectroscopic redshifts (blue), photometric redshift estimates (yellow), and no redshift information (green). Dashed lines denote accuracy assuming no labels are known, and solid lines denote accuracy with labels known for a randomly sampled 10\% of the dataset. }
    \label{fig:accuracy_comparison}
\end{figure}

Figure~\ref{fig:accuracy_comparison} shows the overall accuracy (calculated as \# correctly classified / total \# SNe) for six cases: spectroscopic (blue), photometric (yellow), and no redshift (green) estimates, with no known classifications (dashed lines) and with 10\% known (solid lines), as a function of the true Ibc/Ia ratio in the dataset. For both the spectroscopic and photometric redshift cases, the accuracy meets or exceeds 90\% with no known data, and 87\% with 10\% known data. The best performance occurs for the dataset with a Ibc ratio of 10\%, which has 93.5\% accuracy across the entire dataset for both the spectroscopic and photometric redshift cases, while the accuracy slightly decreases with Ibc ratio. At higher Ibc ratios, there are likely more objects in the overlap region between the two identified Gaussians, leading to slightly more misclassifications. 

In addition to the overall accuracy across the entire dataset, we also calculate the purity and completeness within each class. These are important metrics to fully understand the performance of a classifier, especially in a case such as ours where a class imbalance means that a high overall accuracy could be achieved by just labeling every object as the majority class. For SNe Ia, they are defined as:
\begin{equation}
    \text{Ia purity} = \frac{\# \text{ SNe Ia correctly classified}}{\# \text{ SNe classified as Ia}}
\end{equation}
\begin{equation}
    \text{Ia completeness} = \frac{\# \text{ SNe Ia correctly classified}}{\# \text{ true SNe Ia}}
\end{equation}
while for SNe Ibc, these are calculated with respect to SNe Ibc instead of Ia. If a model classifies everything as SN Ia, it will have 100\% SN Ia completeness, a SN Ia purity equal to the proportion of SNe Ia in the dataset, and zero purity and completeness for SNe Ibc. A perfect classifier would yield 100\% purity and completeness for both classes.

\begin{figure*}
    \epsscale{1.15}
    \centering
    \plotone{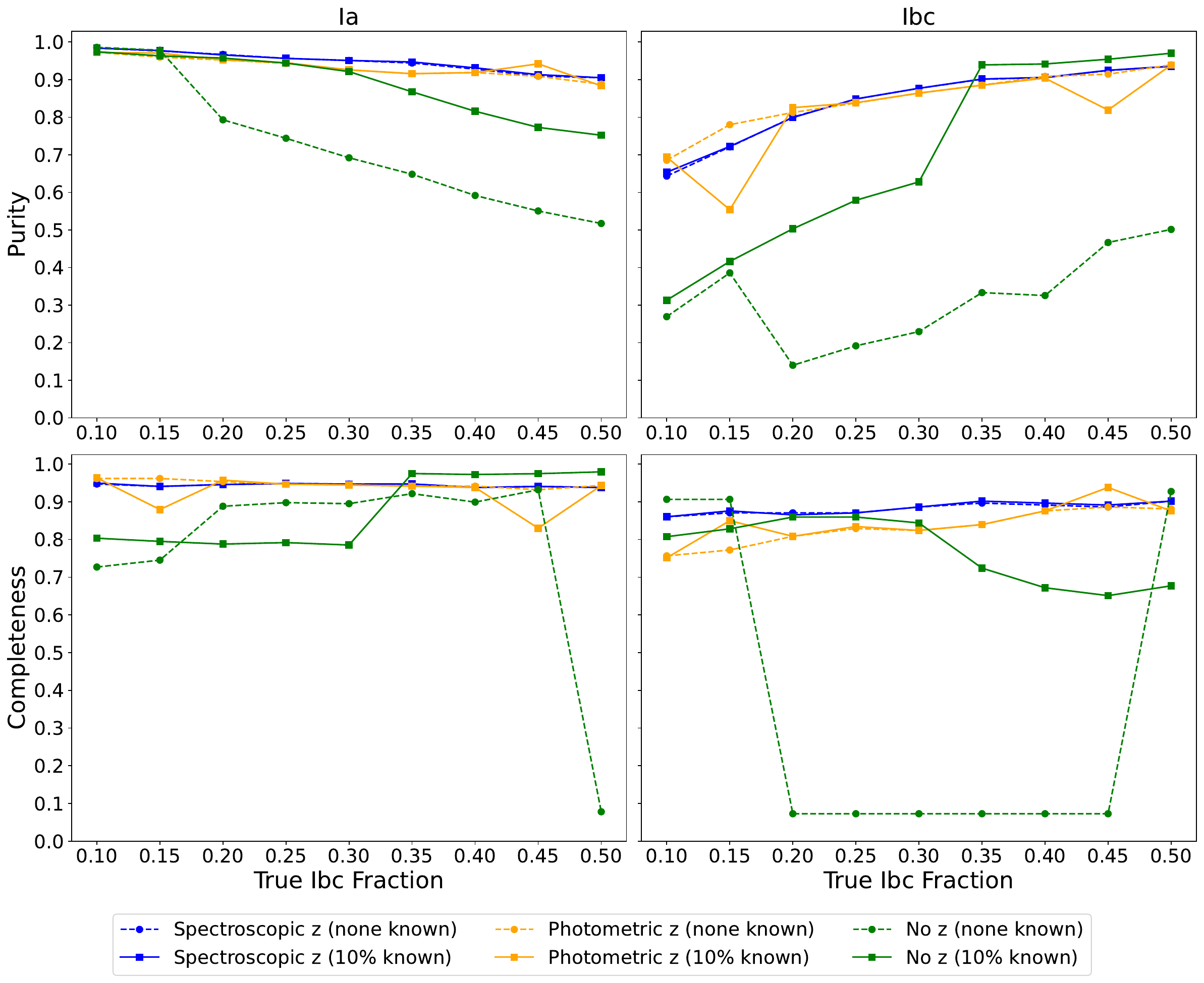}
    \caption{Purity (top row) and completeness (bottom row) for SNe Ia (left column) and SNe Ibc (right column) from our GMM fit based on the $\log_{10} f_{\rm Ni}$ and $\log_{10} v_{\rm ej}$ features as a function of the true SNe Ibc fraction in the population. Light curve fits were done using either spectroscopic redshifts (blue), photometric redshift estimates (yellow), and no redshift information (green). Dashed lines denote accuracy assuming no labels are known, and solid lines denote accuracy with labels known for a randomly sampled 10\% of the dataset.}
    \label{fig:purity_completeness}
\end{figure*}

Figure~\ref{fig:purity_completeness} shows the purity and completeness for both SNe Ia and Ibc for all cases. Again, the spectroscopic and photometric redshift cases are similar, with purity and completeness around 90\% for SNe Ia and around 80\% but increasing with Ibc fraction for SNe Ibc.  The addition of 10\% known classifications does not make a tangible difference in performance. At lower Ibc fractions, our model is slightly biased towards over-classifying SNe as Ibc, yielding a higher purity for SNe Ia than for SNe Ibc. This follows Figure~\ref{fig:mixing_fraction_redshift}, which shows that the fitted mixing fraction is slightly overestimated at lower Ibc fractions, meaning some SNe Ia are being classified as SNe Ibc. Even at a Ibc fraction of 0.2, the purity and completeness for both classes are at or above 80\% for both the photometric and spectroscopic redshifts. 

As seen in Figure~\ref{fig:mixing_fraction_redshift}, the model generally skews slightly towards overestimating the mixing fraction at lower Ibc fractions and underestimating it at higher fractions. This means that at lower fractions, our method classifies more objects as SNe Ibc that the truth, and at higher fractions it is classifying more objects as SNe Ia. This is also reflected in the purity and completeness values in Figure~\ref{fig:purity_completeness}, which shows that values trending downwards with Ibc fraction for SNe Ia and upwards for SNe Ibc.

Across all metrics, the spectroscopic and photometric redshift models have very similar performance, and both perform notably better than that of the model with no redshift information. The inclusion of 10\% known labels does little to improve these models. In the no-redshift case, the additional anchoring of the 10\% known classifications makes a significant difference for the model performance, bringing the no-redshift case much closer to the spectroscopic and photometric redshift cases. The no-redshift case with no known labels has relatively poor performance, with accuracies as low as 50\%, but the addition of known classifications for 10\% of the dataset yields a model with accuracies $\geq$ 80\% for all Ibc fraction values. Appendix~\ref{appendix:fitting_all} provides a discussion of the performance of fitting all parameters from multiple light curve models instead of selecting the most discriminatory features.

\section{Discussion}\label{sec:discussion}

Our results show that indeed, photometry is all you need to classify SNe. As seen in Figures  \ref{fig:mixing_fraction_redshift}, \ref{fig:accuracy_comparison}, and \ref{fig:purity_completeness}, the model using simulated photometric redshift estimates shows only a slight decrease in performance compared to the model using spectroscopic redshifts. The use of the even broad redshift priors from simulated photometric redshifts makes a much more significant difference to the model performance than having 10\% of the population with spectroscopic classifications. At these relatively low redshifts ($z \leq$ 0.2), our 50\% fitting bound to simulate photometric redshifts is quite generous and real photometric redshifts are likely even more precise \citep{kessler_2019}. 

We compare our classification results to Superphot+ \citep{de_soto_superphot_2024}, a state-of-the-art framework trained to rapidly classify SNe based on their ZTF BTS light curves. While Superphot+ is trained to identify five different SN types instead of just two, it still faces the challenging task of having to distinguish between SNe Ibc and Ia. 75\% of the light curves in its training set are SNe Ia and 4\% are SNe Ibc. Using redshift information combined with photometry, it achieves purity and completeness values of 0.98 and 0.93 for SNe Ia and 0.38 and 0.76 for SNe Ibc, respectively. As seen in Figure~\ref{fig:purity_completeness}, our model achieves  significantly better SNe Ibc purities across all Ibc fractions and at larger fractions, slightly lower SN Ia purity but higher SN Ibc completeness. While multi-class classification is a more complex task, this shows that our method is able to achieve comparable results to larger, much more sophisticated methods for SNe Ia/Ibc classification while eliminating the need for retraining for domain shift.

Our proposed classification method is likely successful because we expect SNe Ia and Ibc to have different ranges of Arnett model fit parameters based on their intrinsic physics \citep{sarin_lightcurve_2026}. The higher iron fraction in SNe Ia progenitors leads to higher fitted Nickel fractions in the SNe Ibc model fit. The ejecta mass $M_{\rm ej}$ and velocity $v_{\rm ej}$ can be degenerate, since either could drive an object with a fixed energy. In other words, the Arnett model is a naturally useful feature set.

In the EM algorithm, the $\gamma_{\rm Ia}$ value denotes the probability of a given object belonging to the SN Ia distribution in our GMM \citep{em_algorithm}. In addition to enabling classification, this value can be used as an indication of confidence in the assigned label. SNe with a normalized $\gamma_{\rm Ia}$ close to 1 are more confidently identified as SNe Ia, while SNe with more intermediate $\gamma_{\rm Ia}$ values are likely to be more ambiguous (note that the $\gamma_{\rm Ia}$ is not necessarily calibrated to be interpreted as a true probability). Inspection of the light curves with $\gamma_{\rm Ia}$ close to 0.5 reveal SNe that have a low number of points, or SNe Ibc with slightly extended $r$ band duration compared to the $g$ band, similar to that of SNe Ia.  These values can potentially be used to identify more ambiguous candidates for spectroscopic follow-up.


To ensure that we have the optimal GMM fit parameters to use for classification, we perform 30 repetitions of the GMM fitting, each with a different random initialization, and choose the fit with the highest log-likelihood. In doing this, we are effectively sparsely sampling the likelihood function of $\alpha$, the mixing fraction and attempting to find the maximum likelihood estimate. To quantify the uncertainty on the mixing fraction, this likelihood could be sampled further to characterize its shape and spread. While it is not guaranteed to Gaussian, the standard deviation of the estimates of $\alpha$ would likely provide a reasonable estimate of its uncertainty. 

The diversity of the transients observed with Rubin is much more significant than that of our training dataset, which includes only objects identified as SNe Ia or Ibc. Our model assumes that there are only two possible classes and does not include a path for identifying additional classes if they are present. The model could be extended to fit three or more components to the dataset instead of two. It would be possible to run multiple GMM fits with different numbers of components and compare the log-likelihoods to determine the optimal number of Gaussian components (and thus SN classes) that could fit the data. However, there is no guarantee that the $f_{\rm Ni}$ and $v_{\rm ej}$ would reflect the distinctions between SNe Ia, Ibc and additional classes, and the decision of which light curve parameters should be fit with the GMM should be revisited.

Now that our method has been validated and the parameters to be used in the GMM are chosen, this method is robust to domain shift -- transitioning from one survey to another with unique redshift ranges, mixing fractions, etc. We contrast this to ``traditional" supervised classifiers, which typically have to be retrained in order to be applied to light curves from surveys not in the training set. This is not a problem for our method, as we are performing our GMM fit and classification in physical feature space using parameters from the light curve fit. Data from a different survey can be be fit with the same model using {\tt sncosmo} and the same downstream method can be applied. We also do not require spectroscopy or any known labels, meaning that our method can be applied to virtually any dataset of SN light curves, including unlabeled ones, assuming the majority class is known.

Beyond classification, an obvious application of our method is for cosmological parameter estimation. Even a small fraction of SNe Ibc contamination in a dataset of SNe Ia can bias the resulting estimates of distance moduli and thus cosmological parameters such as the Hubble constant  \citep{beams, freaza_modeling_2026}. \cite{beams} show that once the contamination in the dataset can be estimated, it can be marginalized over to produce a more accurate parameter estimate. Our method provides an easy way to identify potential contamination and create a purer sample for cosmological applications.

There are several potential drawbacks of our method that should be considered. While our method is straightforward and light curves can be processed in parallel, each object must be fit with a chosen light curve model (in this case, Arnett-like). Such fitting can take several seconds to minutes per light curve, potentially resulting in longer computational times compared to faster classification methods like applying a forward pass of a pre-trained neural network. If each fit takes 10 CPU seconds, then it would take approximately 120 CPU days to fit one million transients observed by LSST if they are fit in series. The computational cost mostly depends on the number of light curve models being fit, and can be drastically reduced by fitting objects in parallel. 

Our method could also be accelerated using simulation-based inference, where a neural network could be trained on light curve fits and then applied to predict fit parameters for light curves without actually having to fit the models \citep{villar2022amortized,vidal2025hierarchical}. These predicted fit parameters could then be fit with the GMM as before. This would provide significant computational speedup, as the model fitting is the most time-intensive step of the method; however, the neural network would likely have to be trained to predict fit parameters only for light curves from a specific survey. This would require re-training for applications to a different dataset, which would negate the domain-agnostic aspect of our method.

We impose extremely loose bounds on the light curve fit parameters, and allow them to reach nonphysical values to produce statistically reasonable fits. While this improves the discriminatory power of our model, it renders some of the actual fit parameters as physically meaningless, and they should not be used to learn about the intrinsic physics of an object. To characterize an object, additional fits should be run with fit parameters properly constrained to only physical values. Alternatively, the unphysical parameter values likely are assigned to objects that are not physically driven by said model. A cutoff value of $\gamma_{\rm Ia}$ could be imposed to identify ``real" SNe Ibc and eliminate those that have unphysical parameter values.

\section{Conclusions}\label{sec:conclusions}
Here we present a straightforward method using GMMs to classify a large population of SNe based only on their photometric light curves. We focus on SNe Ia and Ibc, as their light curves are most often misclassified as one another. We use ZTF forced photometry of SNe Ia and Ibc, and we fit all of the light curves using the Arnett model \citep{chatzopoulos_generalized_2012} in {\tt sncosmo} \citep{barbary_2025_15019859}. Finally, we fit a 2D GMM to the distributions of $\log_{\rm 10} f_{\rm Ni}$ and $\log_{\rm 10} v_{\rm ej}$ to classify each object based on the distribution it falls into. Our best model yields a classification accuracy $\geq 90\%$ across all Ibc/Ia ratios. We also quantify the purity and completeness of each model for SNe Ia and Ibc and find that including 10\% known labels makes a significant difference for the no-redshift case but not for the spectroscopic or photometric cases. Including rough estimates of photometric redshifts yields results of similar performance to including the real spectroscopic redshifts for each object.


This work shows that observing transients only with photometry provides all of the information necessary to reliably classify them as SNe Ia or Ibc. Our method is robust to domain shift, as it performs classification in the physical parameter space instead of the observed feature space, and it can be used across surveys and observational parameters. This method has significant implications for combining transient datasets across observatories, and for SN classification in large photometric surveys like LSST, where the majority of objects will not be followed up with spectroscopy.

\begin{acknowledgments}
The analysis presented in this work includes code written using GitHub CoPilot and Claude Code. All code for this paper can be found at \url{https://github.com/asmuzsoy/beams_ext}.

ASMU was supported by a National Science Foundation Graduate Research Fellowship and would like to thank Karthik Yadavalli and Yize Dong for helpful discussions.  The Villar Astro Time Lab acknowledges support through the David and Lucile Packard Foundation, the Research Corporation for Scientific Advancement (through a Cottrell Fellowship) and the National Science Foundation under AST-2433718, AST-2407922 and AST-2406110. 

This work is supported by the National Science Foundation under Cooperative Agreement PHY-2019786 (The NSF AI Institute for Artificial Intelligence and Fundamental Interactions, \url{http://iaifi.org/}).
\end{acknowledgments}

\begin{contribution}

ASMU performed the analysis and wrote the paper, VAV served as the primary advisor for this work.


\end{contribution}

%
\facilities{PO:1.2m (ZTF)} 
\software{{\tt astropy} \citep{2013A&A...558A..33A,2018AJ....156..123A,2022ApJ...935..167A}, {\tt sncosmo} \citep{barbary_2025_15019859}, {\tt matplotlib} \citep{Hunter:2007}, {\tt scipy} \citep{2020SciPy-NMeth}, {\tt numpy} \citep{harris2020array}, {\tt scikit-learn} \citep{scikit-learn}}


\appendix

\section{Selecting GMM fit parameters}\label{appendix:fitting_all}
\subsection{Details of SN Ia model}
In finding the best parameters to separate SNe Ia and Ibc with our GMM, we also explored parameters from a SN Ia model. We use the SALT2 model \citep{guy_salt2_2007}, which expresses the light curve using a set of empirical parameters. It models the flux $F$ in terms of phase $p$ and 
wavelength $\lambda$ for a given SN as:
\begin{equation}
    F(p, \lambda) = x_0 \times \left[M_0(p, \lambda) + x_1M_1(p, \lambda) + ...\right]
    \times \exp\left[ c_sCL(\lambda)\right]
\end{equation}
where $M_0$ and $M_1$ are data-driven spectral templates constructed from a set of real SN Ia light curves and $CL$ is the color correction law as a function of wavelength. The parameters $x_0$, $x_1$, and $c_s$ are coefficients on these population-level templates that are fit for each individual SN.

\begin{figure*}
    \epsscale{1.2}
    \centering
    \plotone{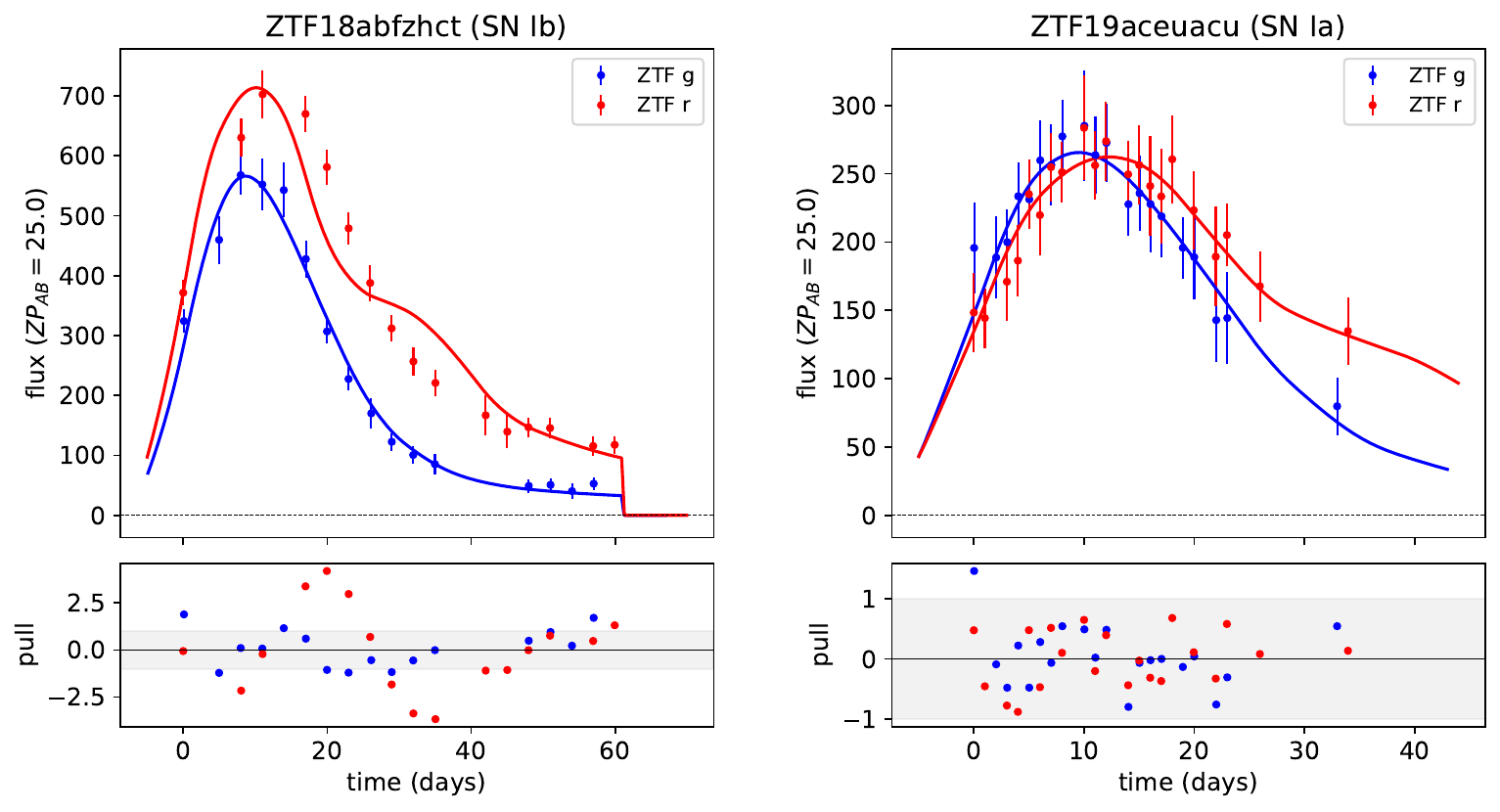}
    \caption{Example fits of true SN Ia and Ibc light curves with the SALT2 SN Ia light curve model. Points indicate observed data points while solid lines indicate model fits. Bottom panels indicate the ``pull", (data - model)/error. Shaded regions indicate a pull between -1 and 1.}
    \label{fig:example_ia_model_fit}
\end{figure*}

We use the built-in {\tt salt2-extended} model in {\tt sncosmo} \citep{barbary_2025_15019859} to fit the light curves using the fitting bounds in Table~\ref{tab:ia_params}. Figure~\ref{fig:example_ia_model_fit} shows example fits for a true SN Ia and Ib with this model. The SN Ia $t_0$ parameter is meant to be the time of peak flux and we constrain it to be greater than 0, i.e. in the actual observed light curve.

\begin{table}[htbp]
\centering
\caption{SN Ia light curve model parameters and fitting bounds.}
\label{tab:ia_params}
\begin{tabular}{lcc}
\hline
Parameter & Lower Bound & Upper Bound \\
\hline
 $c$ & -$\infty$ & $\infty$ \\
 $x_0$  & 0 & 0.1 \\
$x_1$ & -5 & 50 \\
$t_0$ (days) & 0 & 100 \\

\hline
\end{tabular}
\end{table}

\subsection{Fitting all parameters}

When using more than two or three parameters to fit a GMM, inferring a full-rank covariance matrix quickly becomes infeasible with our simple EM algorithm. Instead, when fitting all parameters from both models, we assume a diagonal covariance matrix, fitting the properties of the Gaussian components independently but keeping the mixing fraction $\alpha$ shared among all parameters. This approach allows us to use information from all parameters to constrain the mixing fraction while still maintaining computational efficiency, but it does impose the additional assumption of independence between parameters.

\begin{figure*}
    \epsscale{1.15}
    \centering
    \plotone{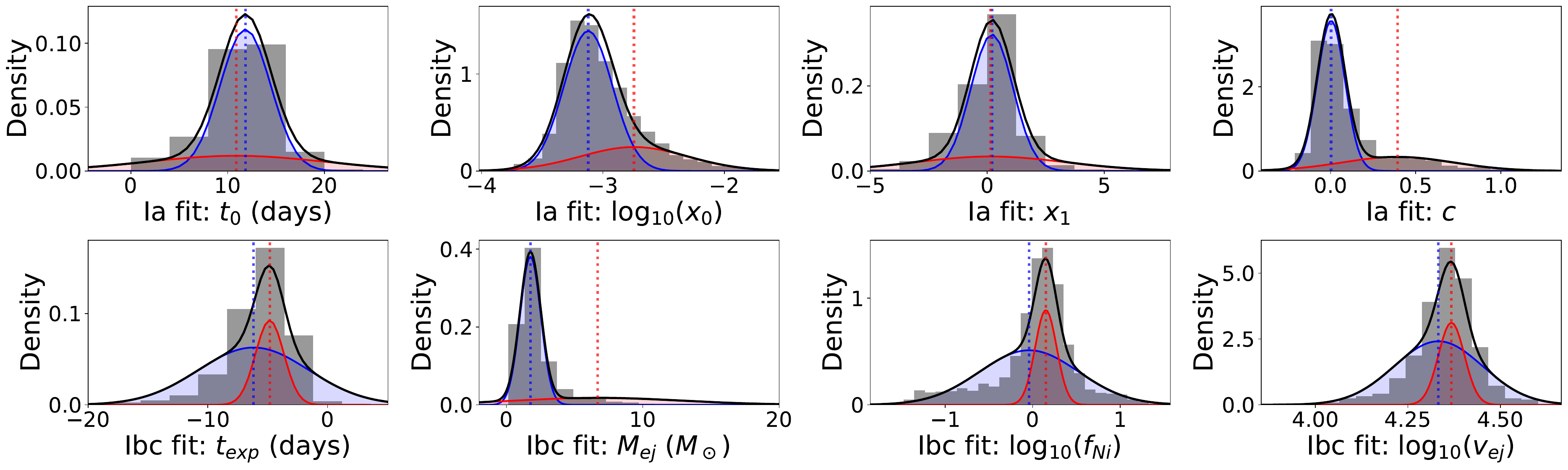}
    \caption{The results of the GMM fitting with the four parameters from the SALT2 SN Ia model and the ``Arnett" SN Ia model. The gray histograms denote all samples, the blue curve denotes the SN Ia component, the red curve denotes the SN Ibc component, and the black curve denotes the mixture of the two. Dashed lines denote the median for each distribution. The true Ibc fraction is 0.1, and fit mixing fraction is 0.275.}
    \label{fig:gmm_results}
\end{figure*}

Figure~\ref{fig:gmm_results} shows the results of the GMM fit on all eight available fit parameters using spectroscopic redshifts in the light curve fits on our entire dataset, which has a Ibc ratio of 0.1. Many of the parameter distributions look like a single Gaussian or cannot be reliably split into the two populations. This model fits the mixing fraction as 0.275 and achieves an overall accuracy of 91\%. It achieves purity and completeness values of 0.91 and 0.99 for SNe Ia and 0.7 and 0.12 for SNe Ibc. By classifying most objects as the majority class (SN Ia), the model has overall poor performance for SNe Ibc.

In practice, this suggests including more fit parameters does not necessarily provide additional constraining power. If each parameter is equally weighted in the classification, then parameters that are not very discriminating (i.e. do not separate cleanly into the two populations) dilute the signal of those that are. This could be mitigated by assigning weights to each parameter based on how informative it is; however, this requires you to use the known labels (i.e., have some spectroscopic training set) to determine how important a particular parameter is. 

\section{Example GMM fits}
\label{appendix:example_fits}
\begin{figure*}
    \epsscale{1.15}
    \centering
    \plotone{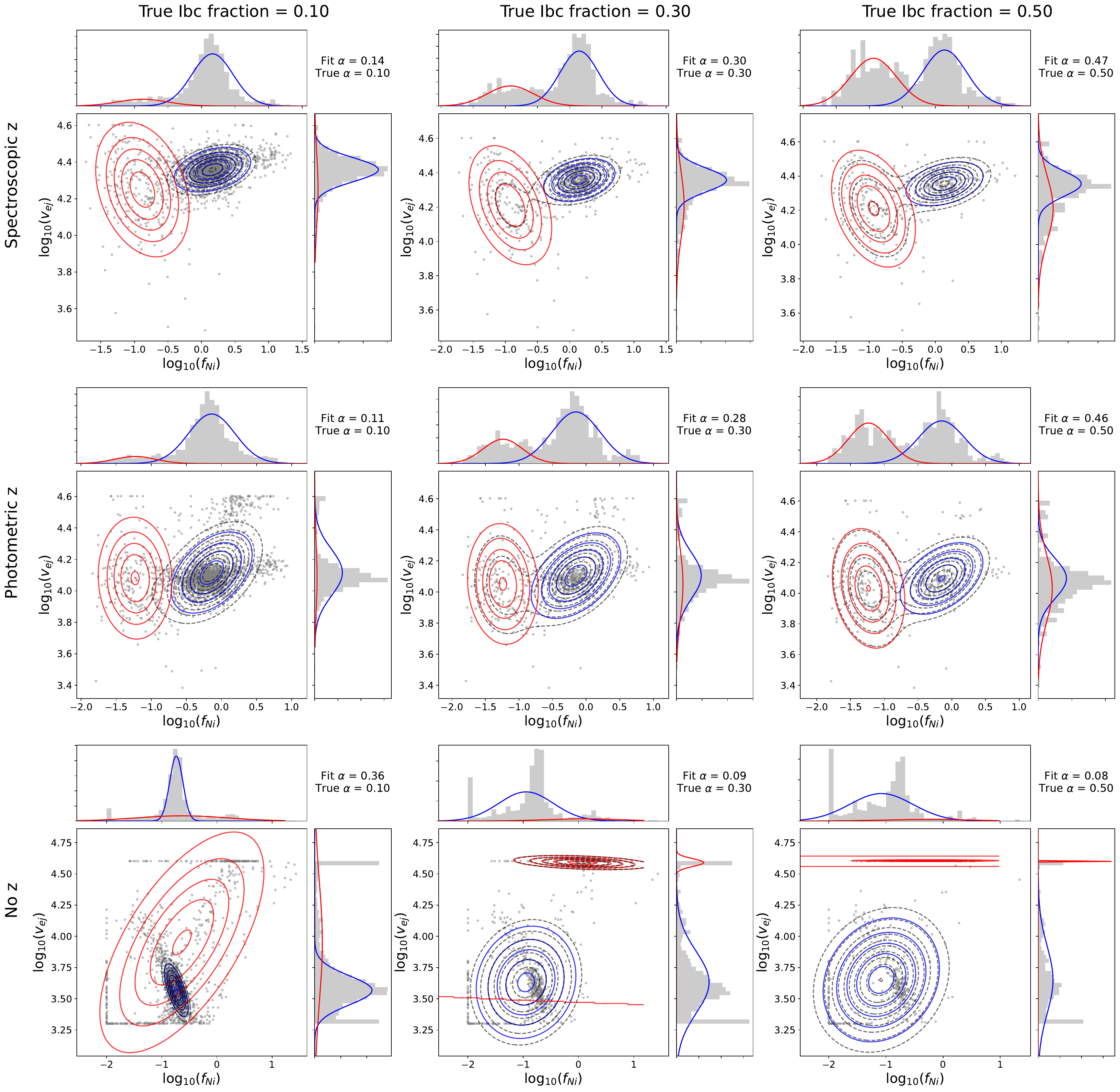}
    \caption{Example fits of our 2D GMM to $\log_{10} f_{\rm Ni}$ and $\log_{10} v_{\rm ej}$. Each gray point is a SN in our sample, and the blue and red curves correspond to the fitted SNe Ia and SNe Ibc components, respectively. Marginal histograms are plotted in gray in the top and right panels of each subplot, and the marginal distributions for SNe Ia and Ibc, scaled by their respective mixing fractions, are plotted in blue and red. The left column of subplots shows fits for a population with a true Ibc fraction of 0.1, the middle column for 0.3 and the right column for 0.5. The top row shows fits using spectroscopic redshifts, the middle row using simulated photometric redshifts, and the bottom row using no redshift information. }
    \label{fig:gmm_grid_none_known}
\end{figure*}

\begin{figure*}\label{fig:sample_fits_known}
    \epsscale{1.15}
    \centering
    \plotone{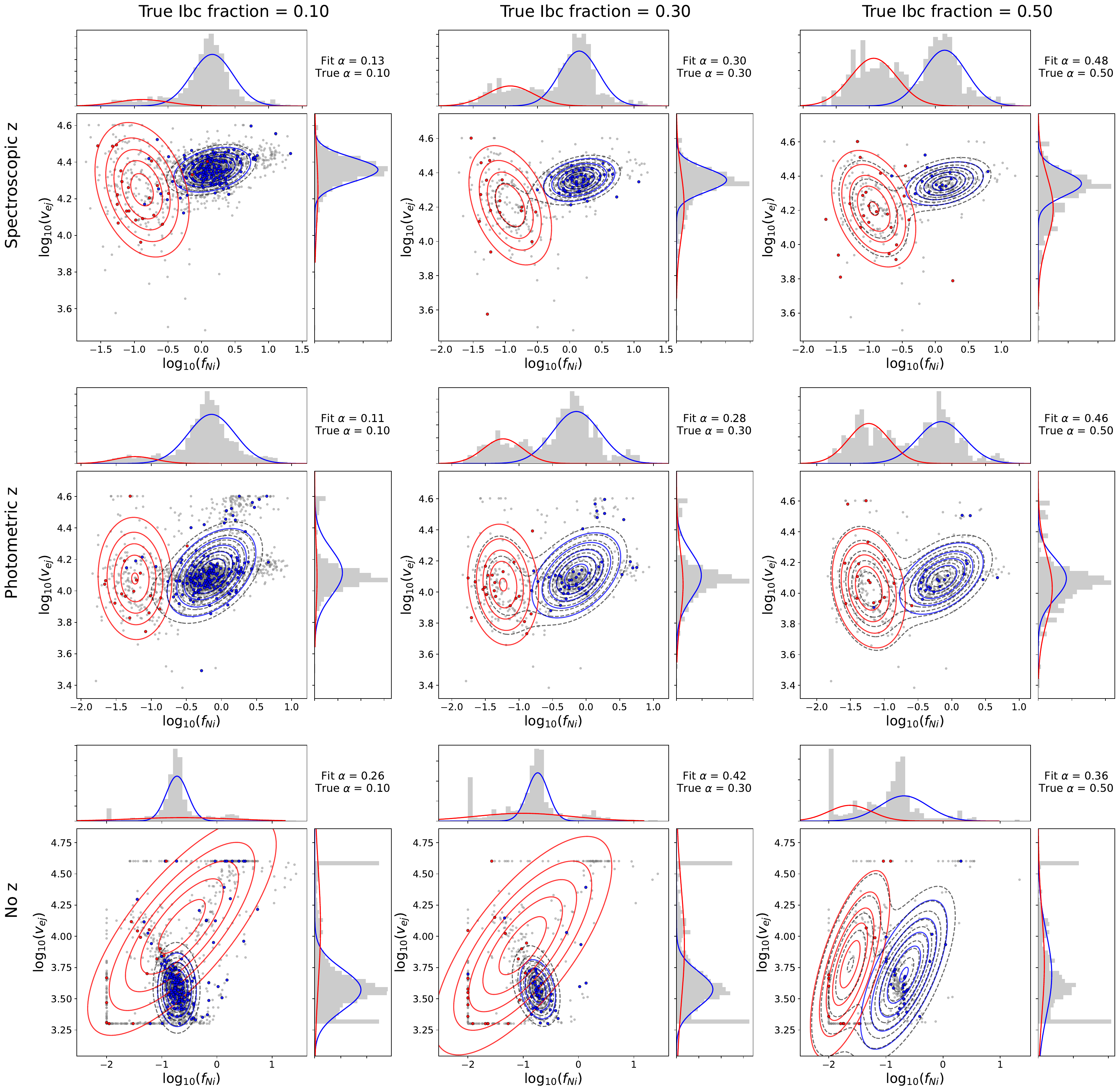}
    \caption{Example fits of our 2D GMM to $\log_{10} f_{\rm Ni}$ and $\log_{10} v_{\rm ej}$. Each gray point is a SN in our sample, and the blue and red curves correspond to the fitted SNe Ia and SNe Ibc components, respectively. Marginal histograms are plotted in gray in the top and right panels of each subplot, and the marginal distributions for SNe Ia and Ibc, scaled by their respective mixing fractions, are plotted in blue and red. The left column of subplots shows fits for a population with a true Ibc fraction of 0.1, the middle column for 0.3 and the right column for 0.5. The top row shows fits using spectroscopic redshifts, the middle row using simulated photometric redshifts, and the bottom row using no redshift information. Colored point denote ``known" points, with red denoting SN Ibc and blue denoting SN Ia. }
    \label{fig:gmm_grid_10_known}
\end{figure*}

Here we show the 2D distribution and 1D marginal distributions of full 2D GMM fits using $\log_{10} f_{\rm Ni}$ and $\log_{10} v_{\rm ej}$. Fits are shown for the none known (Figure~\ref{fig:gmm_grid_none_known}) and 10\% known cases (Figure~\ref{fig:gmm_grid_10_known}), for populations with Ibc fractions of 0.1, 0.3, and 0.5 with spectroscopic, photometric, or no redshift estimates.. 

\bibliography{sample701}{}
\bibliographystyle{aasjournalv7}



\end{document}